\documentclass[final]{svjour2}
\usepackage{graphicx}
\usepackage{rotating}
\usepackage{amsmath}
\usepackage{amssymb}
\usepackage{mathptmx}
\usepackage[numbers,sort&compress]{natbib}
\makeatletter
\journalname{Journal of Low Temperature Physics}

\begin{document}

\newcommand{\hdblarrow}{H\makebox[0.9ex][l]{$\downdownarrows$}-}
\title{Quantum phases and collective excitations of a spin-orbit-coupled Bose-Einstein condensate in a one-dimensional optical lattice}

\author{G. I. Martone}

\institute{LPTMS, CNRS, Univ. Paris-Sud, Universit\'{e} Paris-Saclay, 91405 Orsay, France\\
\email{giovanni.martone@u-psud.fr}}

\maketitle

\begin{abstract}

The ground state of a spin-orbit-coupled Bose gas in a one-dimensional optical lattice is known to exhibit a mixed regime, where the condensate wave
function is given by a superposition of multiple Bloch-wave components, and an unmixed one, in which the atoms occupy a single Bloch state. The unmixed
regime features two unpolarized Bloch-wave phases, having quasimomentum at the center or at the edge of the first Brillouin zone, and a polarized
Bloch-wave phase at intermediate quasimomenta. By calculating the critical values of the Raman coupling and of the lattice strength at the transitions
among the various phases, we show the existence of a tricritical point where the mixed, the polarized and the edge-quasimomentum phases meet,
and whose appearance is a consequence of the spin-dependent interaction. Furthermore, we evaluate the excitation spectrum in the unmixed regime and we
characterize the behavior of the phonon and the roton modes, pointing out the instabilities occurring when a phase transition is approached.

\keywords{Bose-Einstein condensation, spin-orbit coupling, optical lattice}

\end{abstract}

\section{Introduction}
\label{sec:introduction}
After its first experimental achievement by the NIST group~\cite{Lin2011}, Bose-Einstein condensation in the presence of spin-orbit (SO) coupling
has attracted an enormous interest in the community of ultracold atomic gases. The interplay between the modified single-particle dispersion of a
SO-coupled bosonic gas and the two-body interaction allows the realization of exotic configurations, such as spin-polarized states with finite condensation
momentum and striped phases exhibiting a supersolid character
(see the reviews~\cite{Dalibard2011review,Galitski2013review,Zhou2013review,Goldman2014review,Zhai2015review,Li2015review,Zhang2016review} and references therein).
An even richer physics is obtained in the presence of both SO coupling and an optical lattice; the problem of understanding how a shallow or intermediate lattice
affects the static and dynamic properties of a SO-coupled Bose-Einstein condensate (BEC) has been the subject of several
theoretical~\cite{Larson2010,Sakaguchi2013,Zhang2013,Kartashov2013,Cheng2014,Salerno2015,Li2015,Zhang2015,Poon2015,Chen2016,Martone2016b,Hurst2016} and
experimental~\cite{Hamner2015} works. In particular, the ground state of the BEC can be found either in a mixed regime, where the condensate wave function is given
by a superposition of several Bloch wave components, or in an unmixed one, featuring the macroscopic occupation of a single Bloch wave~\cite{Chen2016,Martone2016b}.
The unmixed regime is further characterized by a magnetic phase transition at increasing lattice strength, whose occurrence can be related to the enhancement of the
density response of the condensate at momenta close to the roton wave vector in the plane-wave phase~\cite{Martone2016b}.

The purpose of this work is to illustrate further signatures of the phase transitions occurring in a SO-coupled BEC in the presence of an optical lattice, both
at the level of the ground state and of the excitation spectrum. The rest of the paper is structured as follows. Section~\ref{sec:single_particle} deals with
the single-particle energy spectrum of a SO-coupled BEC in a one-dimensional optical lattice, revealing the different roles of the Raman coupling and the lattice
potential in determining the properties of the noninteracting ground state. In Sec.~\ref{sec:ground_state} we review the main features of the mean-field ground
state in the presence of two-body contact interaction as discussed in Refs.~\cite{Chen2016,Martone2016b}; we additionally point out some interesting effects
arising because of the spin-dependent part of the interaction, including a tricritical point where the mixed, the polarized and the edge-quasimomentum phases meet.
In Sec.~\ref{sec:dynamics} we investigate the properties of the excitation spectrum of the BEC in the unmixed regime, such as the asymmetric propagation of
the sound waves and the occurrence of a roton mode in the polarized phase, as well as the possibility of observing multiple quenches of the phonon mode
by tuning the lattice strength. We summarize in Sec.~\ref{sec:conclusion}.

\section{Band structure of the noninteracting gas}
\label{sec:single_particle}
We consider a gas of (pseudo)spin-1/2 bosons of mass $m$ with the kind of SO coupling induced by using the NIST experimental scheme~\cite{Lin2011}.
The single-particle Hamiltonian is (we set $\hbar=1$)
\begin{equation}
h_\mathrm{SO} = \frac{1}{2m} [(p_x - k_R \sigma_z)^2 + p_\perp^2]
+ \frac{\Omega_R}{2}\sigma_x + \frac{\delta_R}{2}\sigma_z + V_L(x) \, ,
\label{eq:h_SO}
\end{equation}
where $p_\perp^2 = p_y^2 + p_z^2$, $k_R$ is the momentum transfer from the Raman lasers, $\Omega_R$ is the strength of the Raman coupling, $\delta_R$ is
the detuning from Raman resonance (which we set equal to zero in the rest of this work), and $\sigma_x$ and $\sigma_z$ denote the usual Pauli matrices.
The one-dimensional lattice potential can be written as
\begin{equation}
V_L(x) = s E_L \sin^2(k_L x) \, ,
\label{eq:lattice_pot}
\end{equation}
with $k_L$ the lattice wave vector and $s$ the strength of the potential in units of the lattice recoil energy $E_L = k_L^2 / 2m$. Hamiltonian~(\ref{eq:h_SO})
describes a three-dimensional system in the presence of a one-dimensional SO coupling, characterized by equal Rashba~\cite{Bychkov1984} and
Dresselhaus~\cite{Dresselhaus1955} contributions; the optical lattice is taken along the same direction as the SO coupling, i.e., the $x$ axis.

In the absence of the lattice ($s=0$) Hamiltonian~(\ref{eq:h_SO}) is translationally invariant and thus one can look for eigenstates in the form of plane waves
of the kind $\psi_{\vec{k}}(\vec{r}) = e^{i \vec{k}\cdot\vec{r}} \Phi_{k_x}$, where $\vec{k}$ is the eigenvalue of the momentum and $\Phi_{k_x}$ is a
two-component spinor which depends on $k_x$ because of the SO coupling. The energy spectrum as a function of $\vec{k}$ is made of two branches given by
\begin{equation}
\varepsilon^0_\pm(\vec{k}) = \frac{\vec{k}^2}{2m} + E_R
\pm \left[ \left(\frac{k_R k_x}{m} \right)^2 + \left(\frac{\Omega_R}{2}\right)^2\right]^{1/2} \, ,
\label{eq:h_SO_spectrum_s0}
\end{equation}
where $E_R = k_R^2 / 2m$. The ground state of the noninteracting gas can be determined by looking at the minima of the lower branch $\varepsilon^0_-(\vec{k})$.
One finds that for $\Omega_R < 4 E_R$ the latter exhibits two degenerate minima at two opposite momenta $\vec{k} = \pm k_1 \hat{\vec{e}}_x$, with $\hat{\vec{e}}_x$
the unit vector along the $x$ direction and
\begin{equation}
k_1 = k_R \sqrt{1 - \left(\frac{\Omega_R}{4E_R}\right)^2} \, .
\label{eq:k1}
\end{equation}
For $\Omega_R \geq 4 E_R$ the lower branch has instead a single minimum at $\vec{k} = 0$.

When the periodic potential is turned on the Hamiltonian~(\ref{eq:h_SO}) can be diagonalized by resorting to the formalism of Bloch wave functions~\cite{Ashcroft_Mermin_book},
where the eigenstates are represented in the form of a plane wave times a periodic function having the same periodicity $\pi/k_L$ as the lattice potential~(\ref{eq:lattice_pot}).
By further expanding the periodic function in Fourier series, one can write the Bloch wave as
\begin{equation}
\psi^B_{\vec{k}}(\vec{r}) = e^{i \vec{k}\cdot\vec{r}} \sum_{l\in\mathbb{Z}} \Phi_{k_x + 2lk_L} e^{2 i l k_L x} \, ,
\label{eq:Bloch_wave}
\end{equation}
where now $\vec{k}$ is the quasimomentum and the $\Phi_{k_x + 2lk_L}$'s are the two-component coefficients of the Fourier expansion. 
Notice that at $s=0$ only the $l=0$ term of the summation in Eq.~(\ref{eq:Bloch_wave}) is nonvanishing, and one recovers all the results discussed
above. Henceforth we shall restrict the $x$-component of the quasimomentum to the first Brillouin zone, that is, $-k_L \leq k_x \leq k_L$.

\begin{figure}
\begin{center}
\includegraphics[scale=1]{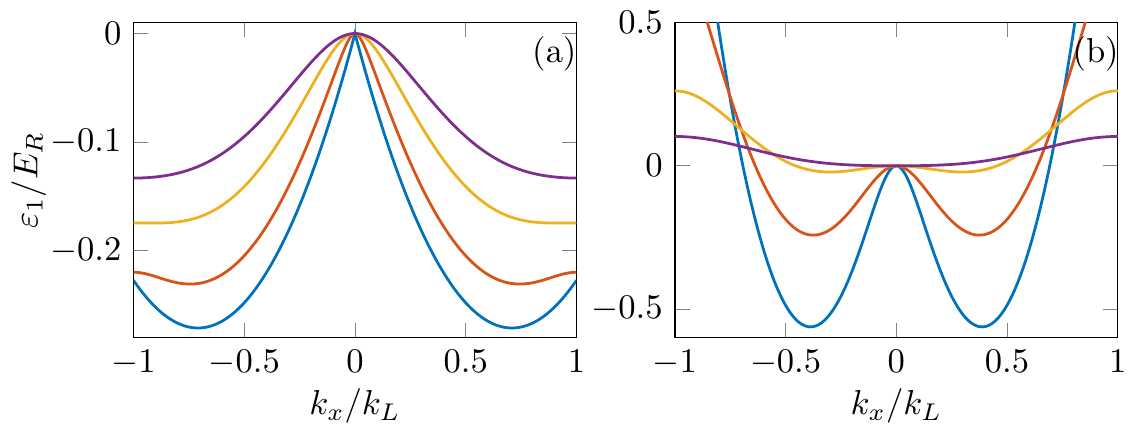}
\end{center}
\caption{(color online) Lowest-lying band $\varepsilon_1$ of the single-particle spectrum at $\Omega_R/E_R = 1.0$ and for several values
of the lattice strength $s$. The parameters for panel (a) are $k_L/k_R = 0.75$, $s = 0$ (blue line), $s = 1.0$ (red line), $s = 2.0$ (yellow line),
$s = 3.0$ (violet line). The parameters for panel (b) are $k_L/k_R = 2.5$, $s = 0$ (blue line), $s = 4.0$ (red line), $s = 8.0$ (yellow line),
$s = 12.0$ (violet line). In order to facilitate the comparison of the shapes of the different curves, in this figure we have redefined the zero
of the energy such that one always has $\varepsilon_1(\vec{k}=0)=0$.}
\label{fig:lowest_band}
\end{figure}

\begin{figure}
\begin{center}
\includegraphics[scale=1]{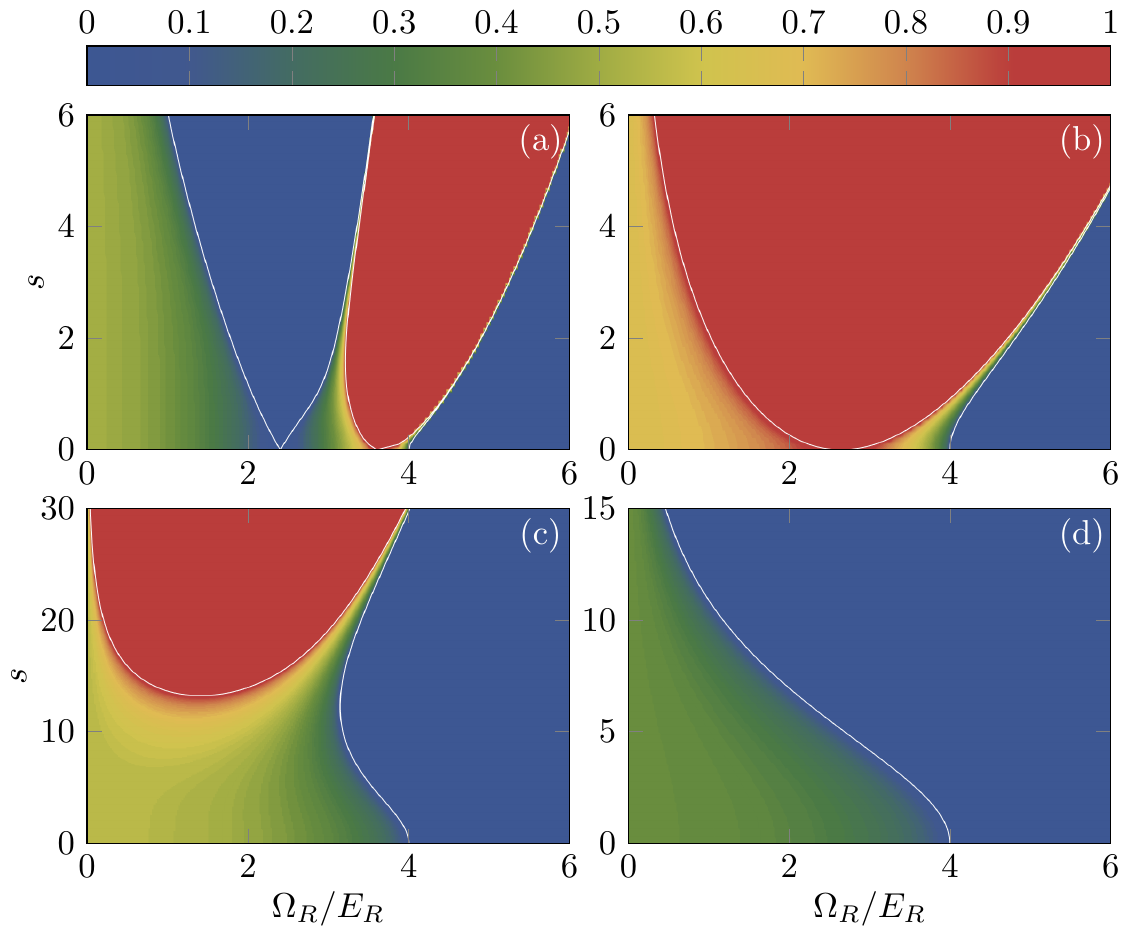}
\end{center}
\caption{(color online) Magnitude of the ground state quasimomentum $k_s/k_L$ as a function of the Raman coupling $\Omega_R$ and the lattice strength $s$
for four different lattice wave vectors: $k_L/k_R = 0.4$ (a), $k_L/k_R = 0.75$ (b), $k_L/k_R = 1.8$ (c), and $k_L/k_R = 2.5$ (d). The white lines separate
the different regions in the diagrams with $k_s/k_L = 0$ (blue), $k_s/k_L = 1$ (red), and $0 < k_s/k_L < 1$ (intermediate colors).}
\label{fig:ks_diag_Omega_sl}
\end{figure}

As expected for a system in the presence of a periodic potential, the energy spectrum exhibits a band structure, and in order to find the ground state one has to look for
the values of $\vec{k}$ at which the lowest-lying band attains its global minima. The locations of such minima and, more generally, the shape of the lowest band of our
SO-coupled BEC are fixed by the interplay between the Raman coupling proportional to $\Omega_R$ and the lattice potential with strength $s$ and wave vector $k_L$. If both
$\Omega_R$ and $s$ are small, the lowest band has two degenerate minima at opposite quasimomenta $\vec{k} = \pm \vec{k}_s = \pm k_s \hat{\vec{e}}_x$ with $k_s \geq 0$.
At large enough $\Omega_R$ the magnitude of the ground-state quasimomentum $k_s$ always shifts towards zero. On the other hand, by increasing the lattice strength at fixed
$\Omega_R$ the minima can move either toward the edge of the first Brillouin zone at $k_x = \pm k_L$ [see Fig.~\ref{fig:lowest_band}(a)] or toward its center at $k_x = 0$
[Fig.~\ref{fig:lowest_band}(b)], the choice between the two being determined by the value of the lattice wave vector $k_L$. In Fig.~\ref{fig:ks_diag_Omega_sl} we plot $k_s$
as a function of $\Omega_R$ and $s$ for different $k_L$'s. Notice that as the lattice strength grows $k_s$ always shifts toward $k_L$ if $1/2 < k_L/k_R \leq 1$
[Fig.~\ref{fig:ks_diag_Omega_sl}(b)] and toward zero if $k_L/k_R \geq 2$ [Fig.~\ref{fig:ks_diag_Omega_sl}(d)], irrespective of the value of the Raman coupling strength;
in the $1 < k_L/k_R < 2$ case [Fig.~\ref{fig:ks_diag_Omega_sl}(c)] $k_s$ always coincides with $k_L$ at sufficiently large $s$, but for intermediate $\Omega_R$ it can first
decrease to zero as the lattice strength is ramped up. A more involved situation occurs when $k_L/k_R \leq 1/2$, as in Fig.~\ref{fig:ks_diag_Omega_sl}(a), since the behavior
of the minima of the lowest-lying band as $s$ is increased is strongly sensitive to the value of $\Omega_R$. We finally notice that, for the ideal Bose gas considered in the
present section, the critical lattice intensity at which the ground-state quasimomentum reaches the center or the edge of the Brillouin zone vanishes whenever $\Omega_R$
is such that the condition $k_1 = n k_L$, with $k_1$ given by Eq.~(\ref{eq:k1}) and $n$ an arbitrary integer number, is satisfied~\cite{Martone2016b}; this happens once in
Fig.~\ref{fig:ks_diag_Omega_sl}(b) and twice in Fig.~\ref{fig:ks_diag_Omega_sl}(a).

\section{Many-body mean-field ground state}
\label{sec:ground_state}
Let us now study how the behavior of the system changes with respect to that of the ideal gas if the particles interact through a two-body contact potential.
For sufficiently weak optical lattices, such that the tunneling strength between adjacent lattice sites is much larger than the on-site interaction energy
between the bosons, quantum fluctuations are expected to play a minor role and the Gross-Pitaevskii (GP) mean-field theory is applicable. Within this approach 
the state of our interacting BEC is described by a two-component wave function $\Psi$ normalized such that $\int_V d\vec{r} \, \Psi^\dagger(\vec{r}) \Psi(\vec{r}) = N$,
where $N$ is the number of particles and $V$ is the volume occupied by the gas. The energy of the system as a functional of $\Psi$ and $\Psi^\dagger$
reads~\cite{Pitaevskii_Stringari_book,Pethick_Smith_book}
\begin{equation}
E[\Psi,\Psi^\dagger] = \int_V d\vec{r} \left\{\Psi^\dagger(\vec{r}) h_{\mathrm{SO}} \Psi(\vec{r}) + \frac{g_{dd}}{2}n^2(\vec{r})
+ \frac{g_{ss}}{2}s_z^2(\vec{r}) + g_{ds} n(\vec{r}) s_z(\vec{r}) \right\} \, ,
\label{eq:E}
\end{equation}
with $n(\vec{r}) = \Psi^\dagger(\vec{r}) \Psi(\vec{r})$ the total density and $s_z(\vec{r}) = \Psi^\dagger(\vec{r}) \sigma_z \Psi(\vec{r})$ the density of the
third spin component. The coupling constants appearing in Eq.~(\ref{eq:E}) correspond to the combinations $g_{dd} = (g_{\uparrow\uparrow}
+g_{\downarrow\downarrow}+2g_{\uparrow\downarrow})/4$, $g_{ss} = (g_{\uparrow\uparrow}+g_{\downarrow\downarrow}-2g_{\uparrow\downarrow})/4$,
and $g_{ds} = (g_{\uparrow\uparrow}-g_{\downarrow\downarrow})/4$ of the interaction strengths in the up-up, down-down and up-down spin channels; the latter
are related to the corresponding $s$-wave scattering lengths via $g_{\sigma\sigma'} = 4\pi a_{\sigma\sigma'}/m$ with $\sigma,\sigma'=\uparrow,\downarrow$.
In the following we shall take $g_{ds}=0$. The wave functions associated to stationary configurations of the system can be obtained by solving the time-independent
GP equation $\delta E/\delta \Psi^\dagger = \mu\Psi$, i.e.,
\begin{equation}
\left[h_\mathrm{SO} + g_{dd}(\Psi^\dagger\Psi) + g_{ss}(\Psi^\dagger\sigma_z\Psi)\sigma_z \right] \Psi = \mu\Psi \, ,
\label{eq:ti_GP}
\end{equation}
where $\mu$ is the chemical potential, whose value is found by imposing that $\Psi$ satisfies the normalization condition discussed above. For any given set of values
of the parameters Eq.~(\ref{eq:ti_GP}) can have multiple different solutions; in particular, the ground state is given by the one yielding the lowest value of the
energy~(\ref{eq:E}).

Let us now consider the thermodynamic limit of the system, which consists of taking $N\to\infty$ and $V\to\infty$ while holding the average particle density
$\bar{n} = N/V$ fixed, and study its zero-temperature phase diagram. The latter can be deduced by calculating the ground state as a function of the parameters
$\Omega_R$, $k_L$, $s$, $\bar{n}$ and $g_{ss}/g_{dd}$ (we use $k_R$ and $E_R$ as the momentum and energy units, respectively). We start by reviewing the results of
Refs.~\cite{Ho2011,Li2012PRL}, which addressed the problem in the absence of the external periodic potential ($s=0$). In the case of antiferromagnetic spin-dependent
coupling, i.e., $g_{ss}>0$, a striped phase with vanishing magnetic polarization $\langle\sigma_z\rangle$ is favored at small Raman coupling $\Omega_R$. In the striped
phase the translational invariance is spontaneously broken, giving rise to periodic modulations in the density profile. The energetic cost of the modulations increases
with the Raman coupling; if $\Omega_R$ exceeds a critical value, which in the $\bar{n} \to 0$ limit is given by the density-independent expression~\cite{Ho2011,Li2012PRL}
\begin{equation}
\Omega_{\mathrm{ST-PW}} = 4 E_r \sqrt{\frac{2 g_{ss}}{g_{dd}+2g_{ss}}} \, ,
\label{eq:Omega_tr_st_pw}
\end{equation}
the system enters a plane-wave phase. The latter features a macroscopic occupation of a state with finite momentum $k_1$ and magnetic polarization per particle
$\langle\sigma_z\rangle / N = k_1/k_R$, where $k_1$ differs from its single-particle value~(\ref{eq:k1}) because of the spin-dependent interaction proportional to
$g_{ss}$~\cite{Li2012PRL}. As in the single-particle case, this configuration is degenerate with the one having opposite momentum $-k_1$ and magnetic polarization
per particle $\langle\sigma_z\rangle / N = - k_1/k_R$. The momentum $k_1$ vanishes when the Raman coupling reaches the value $\Omega_{\mathrm{PW-ZM}}
= 2(2 E_r - g_{ss}\bar{n})$, above which the condensate is in a zero-momentum phase with $\langle\sigma_z\rangle = 0$. We finally mention that in the case of
ferromagnetic spin-dependent coupling ($g_{ss}<0$) the striped phase is energetically unfavored, and the phase diagram contains only the plane-wave and the
zero-momentum phases.

If an optical lattice of the kind~(\ref{eq:lattice_pot}) is present, the ground state can be determined by looking for solutions of the GP equation~(\ref{eq:ti_GP})
of the form
\begin{equation}
\Psi(\vec{r}) = \sqrt{\bar{n}} \sum_{l \in 2\mathbb{Z}+1} C_l \psi_{l \vec{k}_s}^B(\vec{r})
= \sqrt{\bar{n}} \sum_{l \in 2\mathbb{Z}+1} C_l \sum_{l' \in \mathbb{Z}} \Phi_{l l'} e^{i (l k_s + 2 l' k_L) x} \, .
\label{eq:ansatz}
\end{equation}
The Ansatz~(\ref{eq:ansatz}) consists of an infinite sum of Bloch waves of the kind~(\ref{eq:Bloch_wave}) with quasimomenta $\vec{k} = l \vec{k}_s$, each
one entering the superposition with a (generally complex) weight $C_l$, $l$ being an odd integer. As in Eq.~(\ref{eq:Bloch_wave}), in the r.h.s. of Eq.~(\ref{eq:ansatz})
all the Bloch waves $\psi_{l \vec{k}_s}^B$ have been expanded in plane waves, and the two-component spinor coefficients of the expansion $\Phi_{l l'}$ are taken such that
$\sum_{l' \in \mathbb{Z}} \Phi_{l l'}^\dagger \Phi_{l l'} = 1$ for any $l$ (we use the notation $\Phi_{l l'}$ as a shorthand for $\Phi_{lk_s + 2l'k_L}$). The normalization
condition of $\Psi$ then requires that the $C_l$'s satisfy the constraint $\sum_{l \in 2\mathbb{Z}+1} \left|C_l\right|^2 = 1$. Notice that, since the SO coupling and the
optical lattice are only along the $x$ direction, the dependence of the wave function~(\ref{eq:ansatz}) on the transverse coordinates $y$ and $z$ is trivial.

Equation~(\ref{eq:ansatz}) can be justified as follows. If $g_{dd} = g_{ss} = 0$ the nonlinear GP equation~(\ref{eq:ti_GP}) becomes formally identical to the linear
Schr\"{o}dinger equation for a single particle; according to the discussion in Sec.~\ref{sec:single_particle} the solution for the ground state is of the
kind~(\ref{eq:ansatz}) with arbitrary $C_{\pm 1}$, $C_l = 0$ for $|l| > 1$, and $k_s$ and the $\Phi_{\pm1 l'}$'s equal to those calculated within the single-particle
model. On the other hand, in the presence of interactions Eq.~(\ref{eq:ti_GP}) can no longer be solved by a simple superposition of the two Bloch waves with
quasimomenta $\pm \vec{k}_s$ along $x$ because of the nonlinear terms (the only exception is when one of the two coefficient $C_{\pm 1}$ vanishes and $\Psi$ is made
of a single Bloch wave, as in the unmixed regime discussed below). Hence, an infinite number of higher-order contributions with quasimomenta $\pm 3 \vec{k}_s,
\pm 5 \vec{k}_s, \ldots$ need to be included in the condensate wave function to obtain an exact solution of Eq.~(\ref{eq:ti_GP}). Moreover, as we shall see below, interactions
favor specific values of the $C_l$'s, thereby lifting the degeneracy characterizing the single-particle ground state. The mechanism generating the higher-order Bloch waves
in the wave function~(\ref{eq:ansatz}) generalizes the one occurring in the striped phase in the absence of the external lattice~\cite{Li2013} and is strictly related
to the nonlinearity of the GP theory; in particular, the populations $|C_l|^2$ of the Bloch states with $|l|>1$ grow with the density of the gas, while instead they vanish
in the $\bar{n} \to 0$ limit.

We now insert Eq.~(\ref{eq:ansatz}) into Eq.~(\ref{eq:ti_GP}) and we equate the terms on the two sides which oscillate in space with the same wave vector. This yields
an infinite set of equations involving the magnitude of the quasimomentum $k_s$, the weights $C_l$, the components of the $\Phi_{l l'}$'s and the chemical potential $\mu$.
One can numerically solve these equations for a fixed value of $k_s$, keeping into account the above normalization constraints for the $C_l$'s and the $\Phi_{l l'}$'s,
and then study how the solutions vary with $k_s$.\footnote{Notice that, in order to perform the numerical calculation, one has to truncate the two summations in
Eq.~(\ref{eq:ansatz}) to a finite number of terms with $|l| \leq 2 N_s + 1$ and $|l'| \leq N_L$, where $N_s$ and $N_L$ must be chosen large enough such that all the
relevant contributions to the wave function be retained.} The ground state is found by determining the set of values of the above parameters such that the corresponding
wave function $\Psi$, calculated from Eq.~(\ref{eq:ansatz}), minimizes the energy~(\ref{eq:E}). Before moving on we point out that $k_s$, the $\Phi_{l l'}$'s, and
$\mu$ evaluated in the interacting model are generally different from their counterparts in the single-particle picture discussed in Sec.~\ref{sec:single_particle}.

\begin{figure}
\begin{center}
\includegraphics[scale=1]{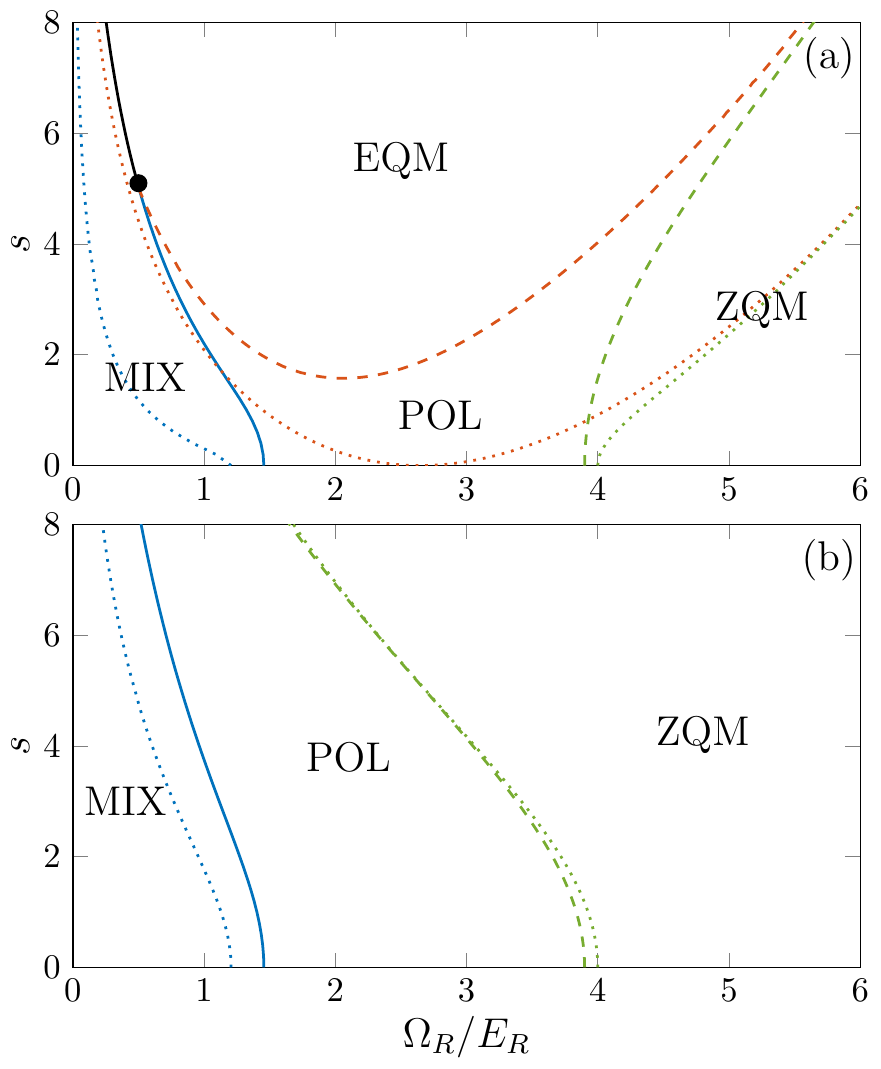}
\end{center}
\caption{(color online) Phase diagram as a function of $\Omega_R$ and $s$ for $k_L/k_R = 0.75$ (a) and $k_L/k_R = 2.5$ (b), $\bar{n} g_{dd}/ E_R = 1.0$,
and $g_{ss}/g_{dd} = 0.05$. The solid curves correspond to the first-order transitions from the mixed to the polarized (blue line) and the edge-quasimomentum (black line)
phases. The red and green dashed curves identify the second-order transitions from the polarized to the edge-quasimomentum and the zero-quasimomentum phases,
respectively. The dotted lines show the behavior of the above transitions at a much smaller density $\bar{n} g_{dd}/ E_R = 10^{-4}$. The black circle in (a) indicates
the position of the tricritical point discussed in the text.}
\label{fig:phase_diag_Omega_sl}
\end{figure}

The results of the above procedure are summarized in Fig.~\ref{fig:phase_diag_Omega_sl}, where we show the phase diagram in the $\Omega_R$-$s$ plane, for fixed
$\bar{n}$ and $g_{ss}/g_{dd}$ (we focus on the most interesting case $g_{ss}>0$). For simplicity we only consider two different values of the lattice wave vector $k_L$, equal
to those of the single-particle diagrams of Figs.~\ref{fig:ks_diag_Omega_sl}(b) and \ref{fig:ks_diag_Omega_sl}(d). At small values of the Raman coupling $\Omega_R$ the system
is in the so-called mixed (MIX) phase~\cite{Chen2016,Martone2016b}, where all the coefficients $C_l$ in the wave function~(\ref{eq:ansatz}) take nonvanishing values and verify
$|C_{-l}| = |C_l|$. In particular, for the values of the parameters used in the present work we always find $|C_{\pm1}|^2 \sim 0.5$ and $|C_l|^2 \ll 1$ for $|l|>1$. Neglecting
the $|l|>1$ terms in Eq.~(\ref{eq:ansatz}) one can visualize this configuration as resulting from the condensation of the atoms into an equal-weighted superposition of two Bloch
states of the kind~(\ref{eq:Bloch_wave}) with quasimomenta $\vec{k} = \pm \vec{k}_s$, corresponding to the two degenerate minima of the single-particle spectrum
(see Sec.~\ref{sec:single_particle}). The mixed phase has vanishing magnetic polarization $\langle\sigma_z\rangle$ and spontaneously breaks the discrete translational
symmetry of the lattice potential~(\ref{eq:lattice_pot}), as its wave function contains the additional oscillation wavelength $\pi/k_s$; the resulting spatial modulation of the density
profile are periodic only if $k_L$ and $k_s$ are commensurate. In the $s \to 0$ limit all the $\Phi_{l l'}$'s with $|l'| \neq 0$ in Eq.~(\ref{eq:ansatz}) vanish and one recovers the
wave function of the striped phase~\cite{Li2013}, with the sole oscillation wave vector $k_s$ left.

At larger Raman couplings $\Omega_R$ the system enters an unmixed regime~\cite{Chen2016,Martone2016b} where all the atoms condense in one of the two single-particle
minima, i.e., they occupy a single Bloch state with quasimomentum equal to $+\vec{k}_s$ or $-\vec{k}_s$. These two configurations are degenerate in energy and correspond
to the values $C_{+1}=1$, $C_{l \neq +1} = 0$ and $C_{-1}=1$, $C_{l \neq -1} = 0$ of the weights in the superposition~(\ref{eq:ansatz}), respectively. Because of
these properties the behavior of the system in the unmixed regime is reminiscent of the single-particle physics discussed in Sec.~\ref{sec:single_particle}. 
In particular, if both $\Omega_R$ and $s$ are sufficiently small the two above degenerate states are physically distinct, being characterized by finite opposite values
of the magnetic polarization $\langle\sigma_z\rangle$, and we say that the condensate is in the polarized (POL) Bloch-wave phase. Instead, if $s$ is small but $\Omega_R$
is large the magnitude of the condensation momentum $k_s$ vanishes giving rise to an unpolarized zero-quasimomentum (ZQM) phase. Notice that in the $s \to 0$ limit
the polarized and the zero-quasimomentum phases approach the plane-wave and zero-momentum states discussed above, respectively.

As already pointed out in Sec.~\ref{sec:single_particle}, the physics at large $s$ depends on the value of the lattice wave vector $k_L$. In the case of
Fig.~\ref{fig:phase_diag_Omega_sl}(a) an increase of the lattice strength causes the condensation quasimomentum $\pm \vec{k}_s$ to move toward the edge of the Brillouin zone,
which is accompanied by a decrease of the magnetic polarization $\langle\sigma_z\rangle$~\cite{Chen2016,Martone2016b}; eventually, when $k_s$ becomes equal to $k_L$,
a new unpolarized configuration appears, which we refer to as the edge-quasimomentum (EQM) phase. In Fig.~\ref{fig:phase_diag_Omega_sl}(b) one has instead that $k_s$
decreases and finally vanishes with increasing $s$, i.e., the condensation quasimomentum is shifted to the center of the Brillouin zone, which yields again the above
zero-quasimomentum phase.

The transitions from the mixed to the polarized and the edge-quasimomentum phases are of first order, and can be revealed by the change in the momentum distribution
of the condensate and in the behavior of the oscillations of the density profile. The mixed-to-polarized transition also entails a sudden jump in the magnetic polarization
$\langle\sigma_z\rangle$. We generally find that the critical Raman coupling at which the ground state of the system leaves the mixed phase is maximum at $s=0$, where
it is given (up to corrections due to the finite density of the system) by Eq.~(\ref{eq:Omega_tr_st_pw}), and decreases with increasing lattice strength. Thus, the lattice
favors the unmixed phases over the mixed one, in agreement with the findings of Ref.~\cite{Hurst2016} which considered SO-coupled BECs of spin 1. On the other hand, the mixed
phase becomes energetically more convenient by taking larger values of the average density $\bar{n}$ and of the ratio $g_{ss}/g_{dd}$. It is worth pointing out that for any
given $s$ the mixed-to-unmixed transition occurs at a finite value of $\Omega_R$ even in the limit of vanishingly small $\bar{n}$, which generalizes the analogous result of
Eq.~(\ref{eq:Omega_tr_st_pw}) holding at zero lattice strength~\cite{Ho2011,Li2012PRL}.

The transitions from the polarized to the zero-quasimomentum and the edge-quasimomentum phases are instead of second order, being characterized by a smooth variation
of the condensation quasimomentum $k_s$ and the magnetic polarization $\langle\sigma_z\rangle$, and they are accompanied by the divergence of the magnetic susceptibility,
as shown in Refs.~\cite{Li2012EPL,Zhang2012,Martone2016b}. In the $\bar{n} \to 0$ limit the transition lines approach their noninteracting counterparts (see
Sec.~\ref{sec:single_particle} and the diagrams of Fig.~\ref{fig:ks_diag_Omega_sl}), whereas at finite density they can differ significantly; in particular, notice that
in the case of Fig.~\ref{fig:phase_diag_Omega_sl}(a) the critical value of the lattice strength $s$ needed to induce the transition from the polarized to the
edge-quasimomentum phase exhibits a minimum as a function of $\Omega_R$, but unlike in the ideal gas model it never vanishes. For a fixed density an increase of the
ratio $g_{ss}/g_{dd}$ favors the unpolarized zero-quasimomentum and edge-quasimomentum phases over the polarized one.

A remarkable feature of the diagram of Fig.~\ref{fig:phase_diag_Omega_sl}(a) is the occurrence of a quantum tricritical point separating the mixed, polarized,
and edge-quasimomentum phases. The existence of this tricritical point is a consequence of the spin-dependent part of the interaction and represents one
of the main results of the present work.

\section{Excitation spectrum. Phonon and roton modes}
\label{sec:dynamics}
The rich phase structure illustrated in Sec.~\ref{sec:ground_state} is expected to give rise to interesting effects at the dynamical level, which can be explored by
resorting to the Bogoliubov theory. For this purpose we switch to the time-dependent GP framework, where the condensate wave function $\Psi$ also depends on
time and evolves according to the time-dependent GP equation~\cite{Pitaevskii_Stringari_book,Pethick_Smith_book}
\begin{equation}
i \frac{\partial \Psi}{\partial t} = \left[h_\mathrm{SO} + g_{dd}(\Psi^\dagger\Psi) + g_{ss}(\Psi^\dagger\sigma_z\Psi)\sigma_z\right] \Psi \, .
\label{eq:td_GP}
\end{equation}
Notice that the time-independent Eq.~(\ref{eq:ti_GP}) is recovered from Eq.~(\ref{eq:td_GP}) by considering stationary solutions of the form $\Psi(\vec{r},t)
= e^{- i \mu t} \Psi_0(\vec{r})$. The wave function describing small oscillations of the system about a given stationary configuration $\Psi_0$ can be written
in the form
\begin{equation}
\Psi(\vec{r},t) = e^{- i \mu t} \left[\Psi_0(\vec{r}) + U(\vec{r}) e^{-i \omega t} + V^*(\vec{r}) e^{i \omega t} \right] \, ,
\label{eq:td_fluctuations}
\end{equation}
where $\omega$ is the oscillation frequency and $U(\vec{r})$, $V(\vec{r})$ are the corresponding two-component small oscillation amplitudes. The latter are taken
to satisfy the normalization condition $\int_V d\vec{r} \left[U^\dagger(\vec{r}) U(\vec{r}) - V^\dagger(\vec{r}) V(\vec{r})\right] = 1$. After inserting
Eq.~(\ref{eq:td_fluctuations}) into Eq.~(\ref{eq:td_GP}), retaining only the linear contributions in $U$ and $V$, and equating the terms proportional to
$e^{-i \omega t}$ and $e^{i \omega t}$ on the two sides, one obtains the Bogoliubov equations
\begin{equation}
\begin{pmatrix}
h_\mathrm{SO} - \mu + h_{UU} & h_{UV} \\
- h_{VU}^* & - (h_\mathrm{SO} - \mu + h_{VV})^*
\end{pmatrix}
\begin{pmatrix}
U \\ V
\end{pmatrix}
=
\omega
\begin{pmatrix}
U \\ V
\end{pmatrix}
\, ,
\label{eq:Bogo_eq}
\end{equation}
where
\begin{align*}
h_{UU} = h_{VV} = {} & {}
g_{dd} (\Psi_0^\dagger\Psi_0 + \Psi_0 \otimes \Psi_0^\dagger)
+ g_{ss} \Big[(\Psi_0^\dagger \sigma_z \Psi_0) \sigma_z + (\sigma_z\Psi_0) \otimes (\sigma_z\Psi_0)^\dagger\Big] \, , \\
h_{UV} = h_{VU} = {} & {}
g_{dd} \Psi_0 \otimes \Psi_0^T + g_{ss}  (\sigma_z\Psi_0) \otimes (\sigma_z\Psi_0)^T \, ,
\end{align*}
and we have used the symbol $\otimes$ to denote the ordinary Kronecker product.

In this work we study the excitation spectrum in the phases belonging to the unmixed regime where the ground-state wave function is a single Bloch wave,
i.e., $\Psi_0(\vec{r}) = \sqrt{\bar{n}} \, \psi_{\pm \vec{k}_s}^B(\vec{r})$ with $\psi_{\pm \vec{k}_s}^B(\vec{r})$ having the form~(\ref{eq:Bloch_wave})
and calculated following the procedure of Sec.~\ref{sec:ground_state}. Consequently, the small amplitudes $U$ and $V$ which solve the Bogoliubov
equations~(\ref{eq:Bogo_eq}) can be taken themselves as Bloch waves,
\begin{equation}
\begin{split}
U_{\ell\vec{q}}(\vec{r}) &= e^{i (\vec{q} \pm \vec{k}_s)\cdot\vec{r}} \sum_{l\in\mathbb{Z}} \tilde{U}_{\ell\vec{q},l} e^{2 i l k_L x} \, , \\
\qquad V_{\ell\vec{q}}(\vec{r}) &= e^{i (\vec{q} \mp \vec{k}_s)\cdot\vec{r}} \sum_{l\in\mathbb{Z}} \tilde{V}_{\ell\vec{q},l} e^{2 i l k_L x} \, ,
\end{split}
\label{eq:Bogo_ampl}
\end{equation}
where $\vec{q}$ is the quasimomentum carried by the excitation and $\tilde{U}_{\ell\vec{q},l}$, $\tilde{V}_{\ell\vec{q},l}$ are two-component expansion
coefficients. The upper and lower sign in Eq.~(\ref{eq:Bogo_ampl}) hold for the state with positive and negative quasimomentum along $x$, respectively.
For a fixed value of $\vec{q}$ one finds infinitely many solutions of Eq.~(\ref{eq:Bogo_eq}) having different frequencies, which give rise to a band
structure in the excitation spectrum; to account for this, in Eq.~(\ref{eq:Bogo_ampl}) we have also introduced the band index $\ell = 1,2,3,\ldots$.
The frequency of the $\ell$-th band as a function of $\vec{q}$ will be consequently denoted by $\omega_\ell(\vec{q})$. The coefficients $\tilde{U}_{\ell\vec{q},l}$,
$\tilde{V}_{\ell\vec{q},l}$ and the frequencies $\omega_\ell(\vec{q})$ can be computed by inserting the expressions~(\ref{eq:Bogo_ampl}) for the Bogoliubov
amplitudes into Eq.~(\ref{eq:Bogo_eq}), equating the plane-wave terms having the same wave vector on the two sides, and solving the resulting eigenvalue
equation.

\begin{figure}
\begin{center}
\includegraphics[scale=1]{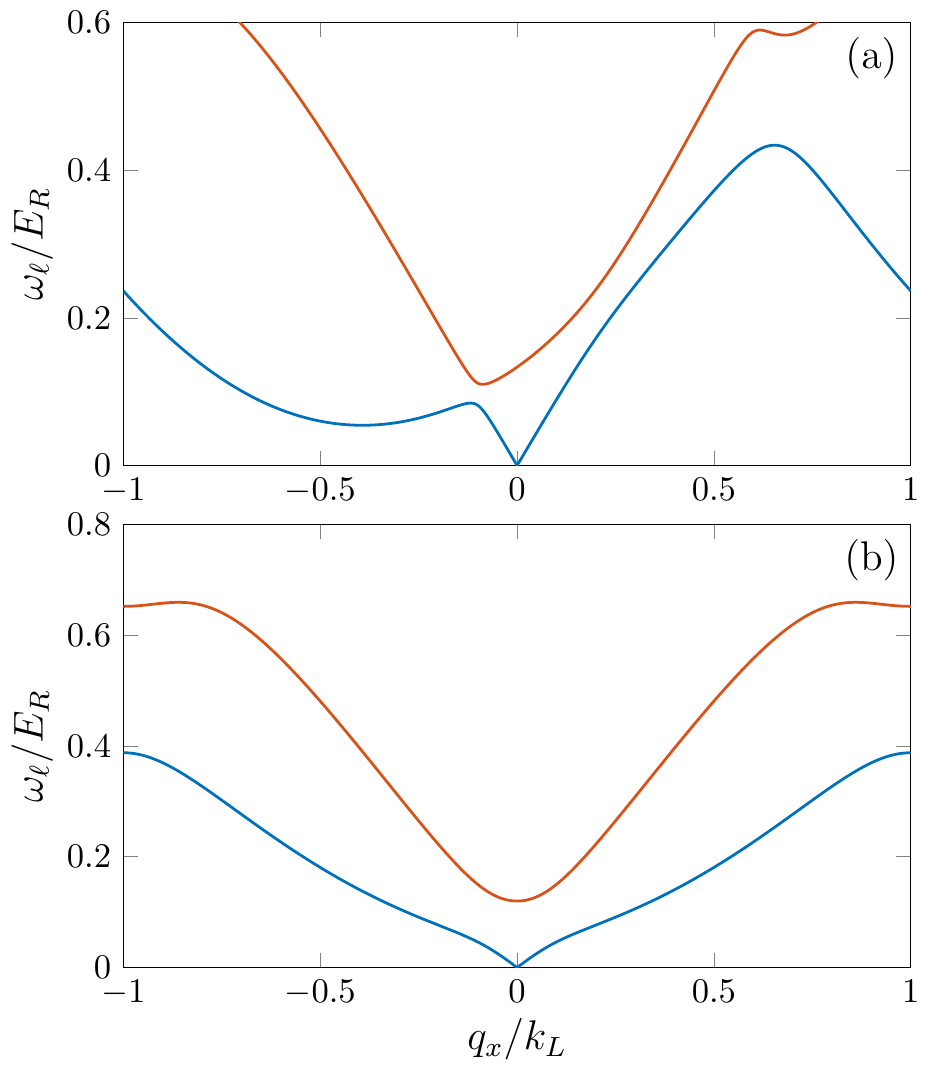}
\end{center}
\caption{(color online) Lowest-lying bands of the excitation spectrum $\omega_\ell$ (blue: $\ell = 1$; red: $\ell = 2$) as functions of the quasimomentum
$q_x$ in (a) the polarized and (b) the edge-quasimomentum phase. The parameters are $\Omega_R/E_R = 2.0$, $k_L/k_R = 0.75$, $s = 0.7$ (a) and $s = 2.0$ (b),
$\bar{n} g_{dd}/ E_R = 1.0$ and $g_{ss}/g_{dd} = 0.05$.}
\label{fig:exc_spectrum}
\end{figure}

The first two bands of the excitation spectra in the polarized and the edge-quasimomentum phases are plotted in Fig.~\ref{fig:exc_spectrum}(a) and
Fig.~\ref{fig:exc_spectrum}(b), respectively. We have considered excitations having quasimomentum $\vec{q}$ along the $x$ axis, for which the modifications
due to the SO coupling are more significant. The main properties of the spectrum in the zero-quasimomentum phase (not shown) are analogous to those of the
edge-quasimomentum phase. Notice that the spectrum in the polarized phase is not symmetric under inversion of $q_x$ into $-q_x$, reflecting the lack of parity
and time reversal symmetry of the ground state; in Fig.~\ref{fig:exc_spectrum}(a) we have chosen to show the results for the polarized state with quasimomentum
$-\vec{k}_s$ directed along the negative $x$ axis.

\begin{figure}
\begin{center}
\includegraphics[scale=1]{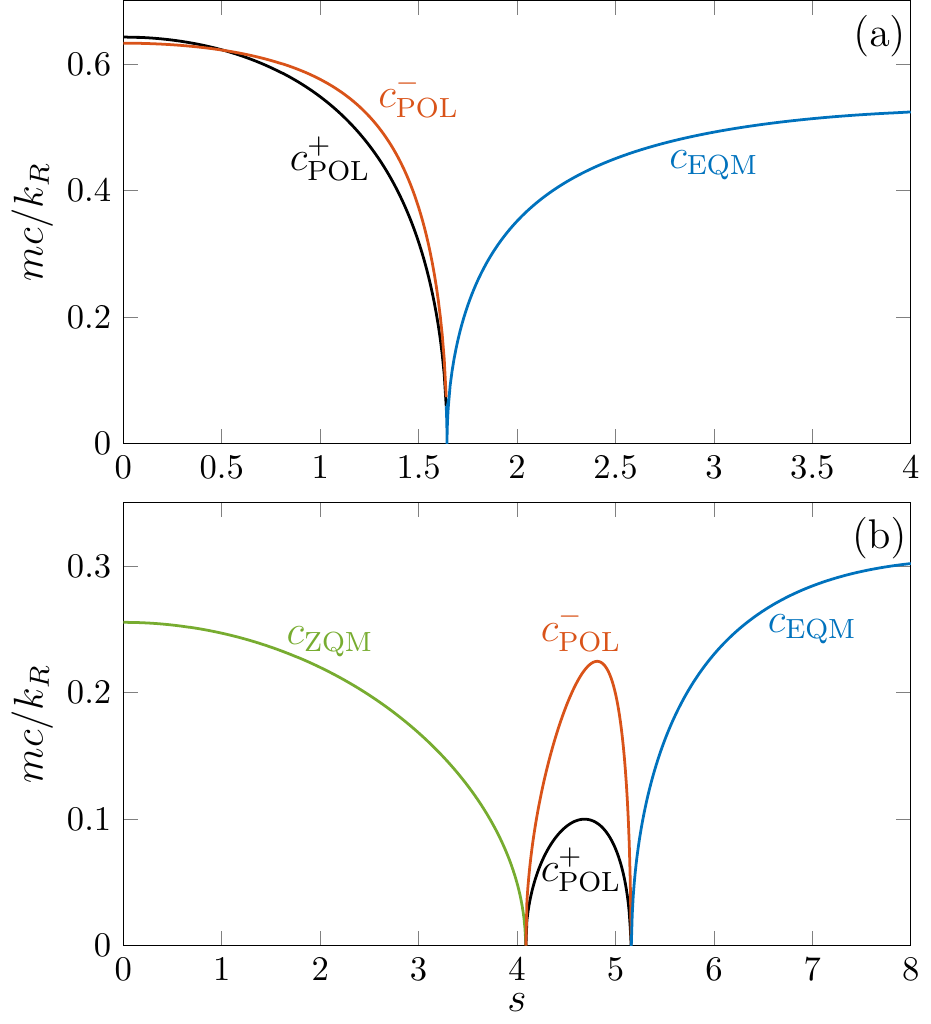}
\end{center}
\caption{(color online) Sound velocity $c$ as a function of the lattice strength $s$ for $k_L/k_R = 0.75$, $\Omega_R/E_R = 0.9$ (a) and
$\Omega_R/E_R = 4.5$ (b). The two different values found in the polarized phase for excitations propagating along the positive and negative
$x$ direction are denoted by $c^+$ and $c^-$, respectively. The other parameters are $\bar{n} g_{dd}/ E_R = 1.0$ and $g_{ss}/g_{dd} = 0.05$.}
\label{fig:sound_vel}
\end{figure}

In both panels of Fig.~\ref{fig:exc_spectrum} one can clearly see that, in the limit of small $q_x$, the lowest-lying band is gapless and exhibits the typical
linear behavior $\omega_{\ell = 1}(q_x) = c q_x$ characterizing the phonon regime of the excitation spectrum of a superfluid, $c$ being the sound velocity.
Additionally, in the polarized Bloch-wave phase the sound velocity is different for excitations with quasimomentum parallel ($q_x > 0$) or antiparallel ($q_x < 0$)
to the $x$ axis (see Fig.~\ref{fig:sound_vel}), similar to what happens in the plane-wave phase at $s=0$~\cite{Martone2012}. The asymmetry of the two sound velocities
is caused by the spin-dependent coupling and disappears if $g_{ss} = 0$. In Fig.~\ref{fig:sound_vel} we plot the sound velocity along $x$ as a function of the lattice
strength $s$, for the same parameters as Fig.~\ref{fig:phase_diag_Omega_sl} and for two fixed values of the Raman coupling $\Omega_R$. One can notice that $c$ undergoes
a quench each time the system crosses one of the second-order transitions between two phases in the unmixed regime. An analogous softening of the phonon mode can be
observed by tuning the Raman coupling across the transition between the plane-wave and the zero-momentum phases in the absence of the lattice~\cite{Martone2012,Ji2015};
however, here the quench is obtained by varying the lattice strength at fixed Raman coupling, and it can occur more than once, as in the case of Fig.~\ref{fig:sound_vel}(b).

We finally point out another peculiarity of the excitation spectrum in the polarized Bloch-wave phase, that is, the occurrence of a roton minimum at finite $q_x$
[see Fig.~\ref{fig:exc_spectrum}(a)], whose energy becomes smaller and smaller as one approaches the transition to the mixed phase. As mentioned in the introduction,
this feature also exists in the plane-wave phase at zero $s$~\cite{Martone2012,Zheng2013,Khamehchi2014,Ji2015} but, as for the quenching of the sound velocity discussed
above, the presence of the lattice provides an alternative mechanism for observing the vanishing of the roton gap. A rotonic behavior has also been found in SO-coupled
BECs with pure Rashba coupling in an optical lattice~\cite{Toniolo2014}.

\section{Conclusion}
\label{sec:conclusion}
The application of a one-dimensional optical lattice to a spin-orbit-coupled Bose-Einstein condensate gives rise to a variety of intriguing phenomena. At the
single-particle level, the increase of the lattice strength can shift the ground state towards an unpolarized configuration with quasimomentum lying either at
the center or at the edge of the first Brillouin zone. When the interactions are taken into account the phase diagram features a mixed regime, where the atoms
occupy a superposition of Bloch states with different quasimomenta, and an unmixed one, where they condense in a single Bloch wave. The various quantum phases
and the corresponding phase transition can be explored by varying the Raman coupling and the lattice strength, and a quantum tricritical point where the mixed,
the polarized and the edge-quasimomentum phases can be identified. At the dynamic level, a typical signature of the phase transitions within the unmixed regime
is represented by the quenching of the velocity of the sound waves propagating along the direction of the spin-orbit coupling; the transition from the polarized
to the mixed phase is instead accompanied by the softening of the roton mode.

From the experimental point of view, the predicted softening of the phonon and roton modes could be measured in currently existing setups with ${}^{87}$Rb atoms,
as it has already been done in spin-orbit-coupled Bose gases without the optical lattice~\cite{Khamehchi2014,Ji2015}. On the other hand, because of the smallness
of the ratio $g_{ss}/g_{dd} \sim 10^{-3}$, the observation of the tricritical point in such experiments would require an exceedingly large value of the lattice strength.
This problem could be solved, for instance, by trapping the atoms in a properly chosen spin-dependent potential~\cite{Martone2014} or by implementing spin-orbit coupling
with the minima of the two lowest-lying bands of an optical superlattice~\cite{Li2016a}, the latter strategy having been recently employed to observe the striped
phase at $s=0$~\cite{Li2016b}.

\begin{acknowledgements}
Useful discussions with T. Ozawa, D. Papoular, N. Pavloff, C. Qu, and S. Stringari are acknowledged.
The research leading to these results has received funding from the European Research Council
under European Community's Seventh Framework Programme (FR7/2007-2013 Grant Agreement No. 341197).
\end{acknowledgements}

\pagebreak

\end{document}